\documentclass{article}
\usepackage[pdftex]{graphicx}

\begin{document}

\title{Quantum communication by means of collapse of the wave function}
\author{Riuji Mochizuki\thanks{E-mail:rjmochi@tdc.ac.jp}\\
Laboratory of Physics, Tokyo Dental College,\\ 2-9-7 Kandasurugadai, Chiyoda-ku, Tokyo 101-0062, Japan }
\maketitle
\begin{abstract}%
We show that quantum communication by means of collapse of the wave function is possible. In this study, {\it quantum communication} means transmission of information itself.  Because of consistency with special relativity, the possibility of the quantum communication leads to another conclusion that the collapse of the wave function must propagate at the speed of light or slower.  
 We show this requirement is consistent with nonlocality in quantum mechanics and also demonstrate that  the Einstein-Podolsky-Rosen experiment does not disprove our conclusion.  
\end{abstract}

\section{Introduction}
As is well-known, violation of the Bell inequality and other Bell-type inequalities\cite{Bell,CH1,FC,CH2,Aspect1,Aspect2,Aspect3,GM,MWZ, Nat} in the quantum world indicates that quantum mechanics cannot be described as a local hidden variable theory, which we call the {\it nonlocality} of quantum mechanics in this study. Because of consistency with special relativity, it is impossible to communicate instantaneously with people who are at a spacelike distance even with the help of this nonlocality. Nevertheless, we should not conclude that we cannot use the transitions of quantum states due to observation as a communication tool; we call these transitions {\it collapse of the wave function}, in accordance with standard texts on quantum mechanics. According to special relativity, it is impossible to communicate by means of the collapse of the wave function if it propagates instantaneously.  Moreover, if the quantum mechanics is unitary, the no-communication theorem prohibits such communication no matter how slowly the information is passed.  Conversely, if the quantum mechanics has a nonunitary process\cite{mochi1}, there is no reason to deny the possibility of slow communication by means of the collapse of the wave function.  Because the proposition that the quantum mechanics is unitary is not certain\cite{Healey}, it is worth examining the possibility of such slow communication. 

Some people may insist that the Einstein--Podolsky--Rosen (EPR) experiment\cite{EPR} demonstrates the instantaneous collapse of the wave function. As discussed below, however, the EPR experiment does not establish the instantaneous collapse of the wave function but the nonlocality of the wave function. We should not confuse them. Equally, communication at a finite speed by means of collapse of the wave function does not conflict with the nonlocality of quantum mechanics.

In this study, we show that quantum communication by means of collapse of the wave function is possible. In this context, {\it quantum communication} does not mean quantum teleportation\cite{telepo1,telepo2,telepo3,telepo4,telepo5} or quantum cryptography\cite{cry1,cry2}, but transmission of information itself.  This study, however, bears some relation to quantum cryptography.  Because of the no-cloning theorem\cite{noclo1,noclo2,noclo3}, the information must change if the communication is tapped by an eavesdropper.  This fact is a hint that communication via quantum entanglement is possible; the eavesdropper can transmit some information to the receiver.  In other words, we may use the no-cloning theorem for communication.

 We consider a thought experiment in which the measurement process is divided into two steps; the first step is a microscopic interaction between the system and the probe of the measuring device, and the second step is the subsequent amplification and output of the result. This thought experiment shows that communication by means of the collapse of the wave function is possible, and no inconsistency with special relativity appears.

This paper is organized as follows. In the second section, we outline our thought experiment. We show in the third section that its inferred results are consistent with special relativity and the nonlocality of quantum mechanics. The EPR--Bohm\cite{Bohm} experiment is discussed in this context in the fourth section. The last section presents our conclusion.  There are some supplementary explanations of entanglement in Appendix.

\section{Thought experiment}
\subsection{Setting}
We consider an experiment in which the spin of an electron $S$ is observed. $|+\rangle$ and $|-\rangle$ are the eigenstates of its spin in the $z$ direction belonging to its eigenvalues $+\hbar/2$ and $-\hbar/2$, respectively. Then, we define 
\begin{equation}
\hat\sigma_z=|+\rangle\langle+|-|-\rangle\langle -|,
\end{equation}
\begin{equation}
\hat\sigma_x=|+\rangle\langle-|+|-\rangle\langle +|.
\end{equation}
$M_z$ and $M_x$ are prepared as macroscopic apparatuses that measure $\hat\sigma_z$ and $\hat \sigma_x$, respectively. The initial state $|r\rangle$ of $S$ is the eigenstate of $\hat\sigma_x$ belonging to its eigenvalue 1:
\begin{equation}
|r\rangle=\frac{1}{\sqrt{2}}\Big(|+\rangle+|-\rangle\Big),\label{eq:appb3}
\end{equation}
\[
\hat\sigma_x|r\rangle=|r\rangle.
\]

$M_z$ includes a microscopic probe that interacts locally with $S$. We define the position operator $\hat\xi_z$ and momentum operator $\hat\pi_z$ of the probe, which satisfy
\begin{equation}
[\hat\pi_z,\ \hat\xi_z]=-i.
\end{equation}
The initial state $|\phi_z\rangle$ of $M_z$ satisfies
\begin{equation}
\langle\phi_z|\hat\xi_z|\phi_z\rangle=\Xi_z,
\end{equation}
\begin{equation}
\langle\phi_z|\hat\pi_z|\phi_z\rangle=0,
\end{equation}
where $\Xi_z$ is a constant.   
A microscopic interaction between $S$ and the probe of $M_z$, which does not include amplification and output of the results in $M_z$, is described by the interaction Hamiltonian $\hat H_z$:
\begin{equation}
\hat H_z\equiv g_z\hat\sigma_z\hat\pi_z,\label{eq:Hz}
\end{equation}
where $g_z$ is a coupling constant.

We define some operators and a state for $M_x$ as we did for $M_z$:
\begin{equation}
[\hat\pi_x,\ \hat\xi_x]=-i,
\end{equation}
\begin{equation}
\langle\phi_x|\hat\xi_z|\phi_x\rangle=\Xi_x,
\end{equation}
\begin{equation}
\langle\phi_x|\hat\pi_x|\phi_x\rangle=0,
\end{equation}
\begin{equation}
\hat H_x\equiv g_x\hat\sigma_x\hat\pi_x.\label{eq:Hx}
\end{equation}
Moreover, the pre-initial state $|\Phi\rangle$ of the combined system of $M_z$, $M_x$, and $S$ is defined as
\begin{equation}
|\Phi\rangle\equiv |\phi_x\rangle|\phi_z\rangle|r\rangle.
\end{equation}

First, $S$ interacts with the probe of $M_z$ for $\Delta t$. If we ignore the time development of $M_z$ itself, the combined state $|\Phi(0)\rangle$ and the density matrix $\hat\rho (0)$ after this interaction are
\begin{equation}
|\Phi(0)\rangle=\exp (-i\hat H_z\Delta t)|\Phi\rangle,\label{eq:state0}
\end{equation}
\begin{equation}
\hat\rho (0)=|\Phi(0)\rangle\langle\Phi(0)|.
\end{equation}
Then, keeping the state unchanged, $M_z$ is separated from $S$ by a distance. We regard this state of the combined system as the initial state at the time $t=0$ in our thought experiment.  Hereafter, we call the operators of $M_z$ and $M_x$ {\it Alice} and {\it Bob}, respectively. 

\begin{figure}
\centering
\includegraphics[width=10cm]{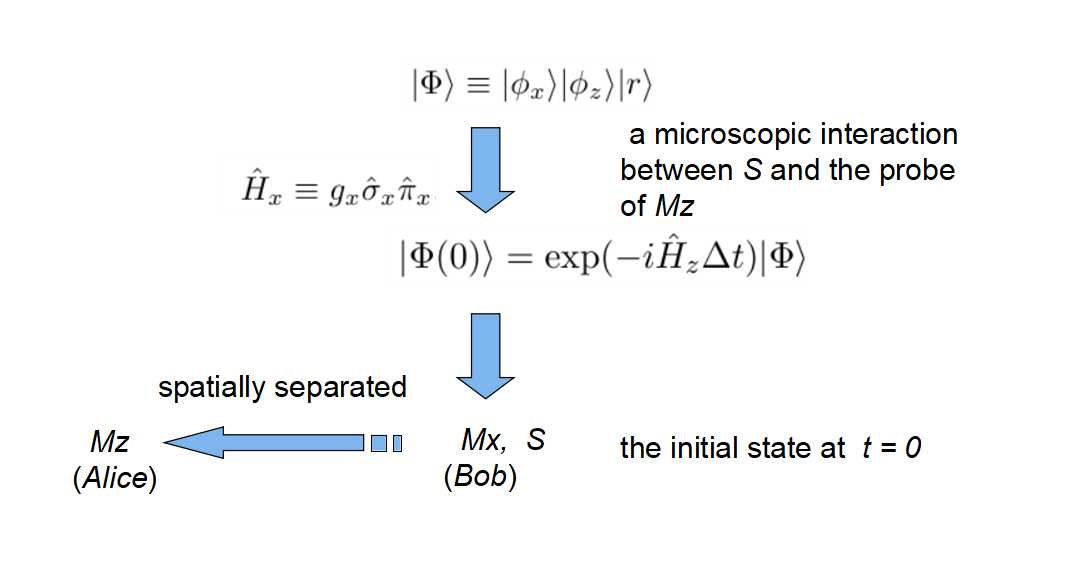} 
\caption{the initial state}
\end{figure}

This experiment consists of the following three operations:\\
1. (the definition of $t_0$) 

The local interaction between $S$ and the probe of $M_x$ for $\Delta t$.  We call the time at which it ends $t_0\ (t_0>0)$.\\
2. (the definition of $t_z$) 

 The measurement processes excluding the local interaction of $S$ with $M_z$, i.e., the amplification and output of the result in $M_z$, which are thought to take little time. We call the time at which they end $t_z\ (t_z>0)$.\\
3. (the definition of $t_x$) 

The measurement processes excluding the local interaction of $S$ with $M_x$, i.e., the amplification and output of the result in $M_x$, which are thought to take little time. We call the time at which they end $t_x\ (t_x\ge t_0)$.

We consider two time orders in which these operations can be performed.

\subsection{$t_z<t_0$}


The thought experiment in this subsection is similar to the three state cascaded Stern--Gerlach experiment\cite{casSG}.
All measurement processes of $M_z$ have finished and the wave function has collapsed before the interaction between $S$ and $M_x$ begins. Therefore, the expectation value $\langle\xi_z\rangle_1$ of $\hat\xi_z$ at $t_z$ is 
\begin{equation}
\begin{array}{rl}
\langle\xi_z\rangle_1&={\rm Tr}\big[\hat\rho(0)\hat\xi_z\big]\\
&=\Xi_z+g_z\Delta t\langle r|\hat\sigma_z|r\rangle\\
&=\Xi_z.\label{eq:41zkekka}
\end{array}
\end{equation}

Then, because of the collapse of the wave function, the density matrix $\hat\rho_1(t_z)$ of $S$ and $M_x$ after $t_z$ becomes
\begin{equation}
\hat\rho_1(t_z)=\frac{1}{2}\Big(|+\rangle\langle +|+|-\rangle\langle -|\Big)\otimes |\phi_x\rangle\langle\phi_x|,\label{eq:deco}
\end{equation}
and the expectation value $\langle\xi_x\rangle_1$ of $\hat\xi_x$ at $t_x$ is
\begin{equation}
\begin{array}{rl}
\langle\xi_x\rangle_1&={\rm Tr}\big[\exp(-i\hat H_x\Delta t)\hat\rho_1(t_z)\exp(+i\hat H_x\Delta t)\hat\xi_x\big]\\
&=\Xi_x+\frac{1}{2}g_x\Delta t\Big(\langle +|\hat\sigma_x|+\rangle+\langle -|\hat\sigma_x|-\rangle\Big)\\
&=\Xi_x.\label{eq:41xkekka}
\end{array}
\end{equation}

\subsection{$t_x<t_z$}
Because no collapse of the wave function has occurred before $t_x$, the state $|\Phi(t_0)\rangle$ of the combined system between $t_0$ and $t_x$ is
\begin{equation}
|\Phi(t_0)\rangle=\exp(-i\hat H_x\Delta t)|\Phi(0)\rangle.\label{eq:phit0}
\end{equation}
Therefore, the expectation value $\langle\xi_x\rangle_2$ of $\hat\xi_x$ at $t_x$ is
\begin{equation}
\begin{array}{rl}
\langle\xi_x\rangle_2&={\rm Tr}\big[|\Phi(t_0)\rangle\langle\Phi(t_0)|\hat\xi_x\big]\\
&=\Xi_x+g_x\Delta t\langle r|\hat\sigma_x|r\rangle\\
&=\Xi_x+g_x\Delta t.\label{eq:44xkekka}
\end{array}
\end{equation}
In the second line of (\ref{eq:44xkekka}), we have neglected the terms of $\mathcal{O}(\hbar)$ and more.  
We associate the thought experiment in this section with the reversible Stern--Gerlach experiment\cite{revSG1,revSG2}, the quantum-eraser experiment\cite{eraser1,eraser2,eraser3} and the delayed-choice experiment\cite{delay1, delay2}.   Alice can cansel the interaction and return the state $|\Phi (0)\rangle$ to $|\Phi\rangle$ before Bob does his work, i.e., before $t_0$. Therefore, we conclude that Alice has no information about the state before $t_z$, and  (\ref{eq:44xkekka}) does not conflict with the no-cloning theorem.

\begin{figure}
\centering
\includegraphics[width=10cm]{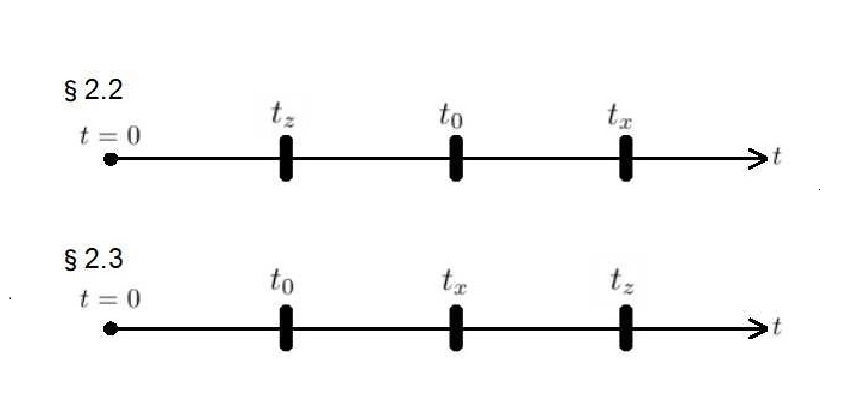}
\caption{time orderings in the 2nd section}
\end{figure}

\subsection{Transmission of information}
Because the two expectation values (\ref{eq:41xkekka}) and (\ref{eq:44xkekka}) are not the same, Alice can change the expectation value of $\hat\xi_x$ and transmit information to Bob. If Alice restarts $M_z$ to amplify and output the results before $t_0$, Bob will obtain the expectation value (\ref{eq:41xkekka}) and learn that {\it the Giants} won. Conversely, if Alice does not restart the measuring process, Bob will obtain (\ref{eq:44xkekka}) and be unhappy to learn that {\it the Giants} lost.
As shown in Appendix B, this result is consistent with the no-communication (no-signaling) theorem.

\section{Special relativity and nonlocality}
\subsection{Requirement of special relativity}
In the previous section, we showed that we can communicate by means of collapse of the wave function. Therefore, the collapse of the wave function should propagate at the speed of light or slower to be consistent with special relativity. On the other hand, the clear violation of the Bell inequality rules out the description of quantum mechanics as a local hidden variable theory. To demonstrate their consistency, we consider other time orderings than those considered in the previous section.  

In the following, we assume that the collapse of the wave function propagates at the speed of light, and hence we use the equal sign (=) for the time ordering of two events that are spacelike separated and $<$ or $>$ for the time ordering of timelike-separated events.

\subsection{$t_0\le t_z<t_x$}
The time ordering between $t=0$ and $t_z$ may change and the state of the combined system just before $t_z$ is (\ref{eq:phit0}) or (\ref{eq:state0}) depending on the frame of reference.   The expectation values of $\hat\xi_z$ for both states are the same: 

\begin{equation}
\langle\xi_z\rangle_3=\Xi_z.\label{eq:42zkekka}
\end{equation}
Therefore, the difference between the frames of reference causes no inconsistency. 

\begin{figure}
\centering
\includegraphics[width=10cm]{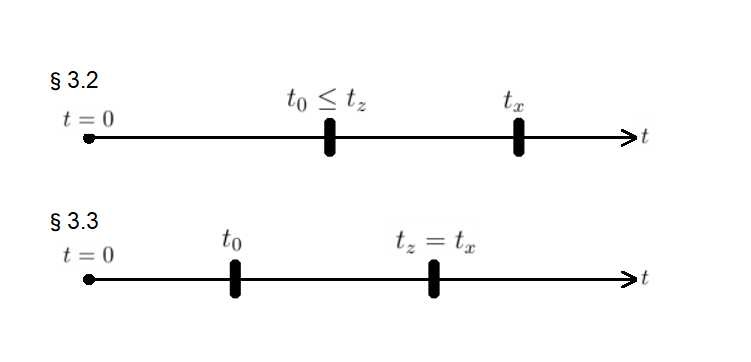}
\caption{time orderings in the 3rd section}
\end{figure}
\subsection{$t_0<t_z=t_x$}

The state of the combined system just before $t_z$ and $t_x$ is (\ref{eq:phit0}), for which both $M_z$ and $M_x$ are operated. Because $\hat\sigma_z$ and $\hat\sigma_x$ do not commute, the Hilbert space to which the state (\ref{eq:phit0}) belongs is not a direct product of the individual Hilbert spaces of $\hat\sigma_z$ and $\hat\sigma_x$. The states of $S$ and $M_z$ are entangled via the interaction whose Hamiltonian is (\ref{eq:Hz}); similarly, the states of $S$ and $M_x$ are entangled via the interaction whose Hamiltonian is (\ref{eq:Hx}). Therefore, the states of $M_z$ and $M_x$ are also entangled, and the measurement processes of $\hat\xi_z$ and $\hat\xi_x$ depend on each other, even though they commute\cite{Arthurs}\cite{moc2}. They are parts of {\it one} measurement process of $\hat\xi_z\hat\xi_x$, whose expectation value $\langle\xi_z\xi_x\rangle_4$ is
\begin{equation}
\begin{array}{rl}
\langle\xi_z\xi_x\rangle_4=&{\rm Tr}\big[|\Phi(t_0)\rangle\langle\Phi(t_0)|\hat\xi_z\xi_x\big]\\
=&\Xi_z\Xi_x+\Xi_zg_x\Delta t\langle r|\hat\sigma_x|r\rangle+\Xi_xg_z\Delta t\langle r|\hat\sigma_z|r\rangle\\
&+\frac{1}{2}g_zg_x(\Delta t)^2\langle r|(\hat\sigma_z\hat\sigma_x+\hat\sigma_x\hat\sigma_z)|r\rangle\\
=&\Xi_z(\Xi_x+g_x\Delta t).\label{eq:43zxkekka}
\end{array}
\end{equation}
Weak value\cite{Aha1, Aha2} is evidence of such entanglement.  As confirmed by many experiments\cite{Res,Lun,Yok}, the measured value of weak measurement agrees with the corresponding weak value.  Although the first measurement and the post-selection are not spacelike-separated but sequential in the weak measurement, similar entanglement of the apparatuses is observed because of the weakness of the first measurement.  See Appendix for the details.

\subsection{Consistency}
The expectation value (\ref{eq:43zxkekka}) is obtained as a necessary consequence of the nonlocality of quantum mechanics and our conclusion in the second section that the collapse of the wave function propagates at the speed of light or slower. For consistency with special relativity, Alice should not be able to know which ordering is correct, $t_z=t_x$ or $t_z<t_x$, and hence the output of $M_z$ described in section 3.3 should not be distinguishable from that in section 3.2. Similarly, Bob should not be able to know which ordering is correct, $t_z=t_x$ or $t_z>t_x$, and hence the output of $M_x$ as described in section 3.3 should not be distinguishable from that of 2.3. As shown in the previous sections, (\ref{eq:44xkekka}), (\ref{eq:42zkekka}), and (\ref{eq:43zxkekka}) are expectation values for the state (\ref{eq:phit0}). Taking account of the commutativity between $\hat\xi_z$ and $\hat\xi_x$ as well, we conclude that it is impossible for Alice or Bob to know about spacelike-separated events despite the nonlocality of quantum mechanics.

In contrast, if we assume that the collapse of the wave function propagates instantaneously, the expectation values of $\hat\xi_x$ for the time ordering $t_z=t_x$ calculated in different frames of reference may be different. That is, this assumption of the instantaneous propagation is inconsistent with special relativity.

\section{Discussion}
In the previous sections, we clarified that communication by means of collapse of the wave function is possible and is consistent with both special relativity and the nonlocality of quantum mechanics. In some textbooks, however, the consistency between special relativity and the nonlocality of quantum mechanics is considered to be guaranteed on the hypothesis that communication by means of quantum correlation is impossible. For example, Redhead claimed this in his famous book\cite{red}, but his proof was inadequate because he ignored entanglement of quantum states. On the other hand, there is an opposite way of explaining things. The impossibility of communication by means of quantum correlation is occasionally regarded not as evidence of consistency but as its necessary consequence. We should be careful not to confuse the nonlocality of quantum mechanics with instantaneous communication by means of quantum correlation\cite{Esp}. Communication by means of collapse of the wave function at the speed of light or slower is consistent with both special relativity and the nonlocality of quantum mechanics, as shown in the previous section.

The main objection to our conclusion may be the assumption that the EPR--Bohm experiment demonstrated instantaneous propagation of the collapse of the wave function. As shown below, however, this objection is wrong. The EPR--Bohm experiment proves only the nonlocality of quantum mechanics. To show this, we consider measurement of the spins in the $z$ direction of an EPR pair of spin 1/2 particles whose initial state $|I\rangle$ is defined as
\begin{equation}
|I\rangle =\frac{1}{\sqrt{2}}\Big( |+\rangle_1|-\rangle_2+|-\rangle_1|+\rangle_2\Big),\label{eq:apb}
\end{equation}
where $|+\rangle_{1(2)}$ and $|-\rangle_{1(2)}$ are the eigenstates of $\hat\sigma_z^{1(2)}$ of particle 1 (2) with eigenvalues $+1$ and $-1$, respectively. 

If the two measurement processes for the two particles are timelike separated, there is no mystery.  The collapse of the wave function propagates from one electron to the other, and the later measurement would be done for the state that has an already-determined spin. 

In contrast, we must carefully observe the experiment if the two measuring processes are spacelike separated.  Firstly, we notice that $|I\rangle$ can also be expressed as an eigenstate of the spin-correlation operator $\hat C$:
\begin{equation}
\begin{array}{rl}
\hat C\equiv&\hat\sigma_z^1\otimes\hat\sigma_z^2\\
=&\big(|+\rangle_1|+\rangle_2\langle +|_2\langle +|_1+|-\rangle_1|-\rangle_2\langle -|_2\langle -|_1\\
&-|+\rangle_1|-\rangle_2\langle -|_2\langle +|_1-|-\rangle_1|+\rangle_2\langle +|_2\langle -|_1\big),
\end{array}
\end{equation}
\begin{equation}
\hat C|I\rangle =-|I\rangle.\label{eq:soukan}
\end{equation}

If the two measurement processes are spacelike separated, our measurement of both spins is regarded as {\it one} measurement of $\hat C$ for the entangled state, because the states of particles 1 and 2 are entangled. In this case, what we obtain is the expectation value of $\hat C$, which is similar to the situation in section 3.3\footnote{In the EPR--Bohm case, the expectation value obtained by each observer can also be interpreted as the independent expectation values of the spin of each particle, because $\hat\sigma_z^1$ and $\hat\sigma_z^2$ commute.}. The spin of each particle cannot be thought to have been determined before the measurement, as confirmed by the violation of the Bell inequality. Nevertheless, the expectation value of $\hat C$ for $|I\rangle$ is $-1$ because $|I\rangle$ is the eigenstate of $\hat C$, which has an eigenvalue of $-1$.  Therefore, accepting the nonlocality of the wave function is sufficient to understand the consistency between our conclusion and the result of the EPR--Bohm experiment.
Note that what we want to demonstrate here is that the EPR-Bohm experiment does not disprove our conclusion, though we cannot explain by means of any local theory {\it why} both of the outcomes are consistent with this fact.

\section{Conclusion}
We showed that quantum communication by means of collapse of the wave function is possible. Therefore, consistency with special relativity requires the collapse of the wave function to propagate at the speed of light or slower. This fact is also consistent with the nonlocality of quantum mechanics.

\section*{Appendix : Entanglement between the states of $M_z$ and $M_x$ in $|\Phi (t_0)\rangle$}

In this appendix, we show that the states of $M_z$ and $M_x$ in (\ref{eq:phit0}) are entangled and weak value is evidence of such entanglement.  The discussion about weak measurement in this Appendix is based on \cite{moc2}.

If $t_0<t_z=t_x$, the state of the combined system just before $t_z$ and $t_x$ is (\ref{eq:phit0}):
\[
\begin{array}{rl}
|\Phi(t_0)\rangle&=\exp(-i\hat H_x\Delta t)|\Phi(0)\rangle\\
&=\exp(-i\hat H_x\Delta t)\exp(-i\hat H_z\Delta t)|\Phi\rangle.
\end{array}
\]
We define the partial density matrix $\hat\rho^{(m)}(t_0)$of the two measuring devices as 
\begin{equation}
\hat \rho^{(m)}(t_0)={\rm Tr}^{(s)}\big[|\Phi(t_0)\rangle\langle\Phi(t_0)|\big],\label{eq:density}
\end{equation}
where ${\rm Tr}^{(s)}$ is the partial trace of the observed system.  

 If the ensembles $\mathcal{M}_z$ of $M_z$ and $\mathcal{M}_x$ of $M_x$ after their unitary interaction with the measured system are both separately obtained by combining all the elements of the sub-ensembles, each of them can be described by its own ket.  Each element of $\mathcal{M}_z$ belongs to one of the sub-ensembles $E_{\alpha},\ \alpha =1,2,\cdots$, described by $|Z_{\alpha}\rangle$ and each element of $\mathcal{M}_x$ belongs to one of the sub-ensembles $E_{\beta},\ \beta =1,2,\cdots$, described by $|X_{\beta}\rangle$ such that the sub-ensemble $\varepsilon_{\alpha,\beta}$ of the combined measuring device, whose elements belong to both $E_{\alpha}$ and $E_{\beta}$, is described by the density matrix
\[
\hat\rho_{\alpha ,\beta}=|X_{\beta}\rangle |Z_{\alpha}\rangle\langle Z_{\alpha}|\langle X_{\beta}|,
\]
and the ensemble of the combined measuring device is described as the weighted sum of $\hat\rho_{\alpha,\beta}$:
\begin{equation}
\hat\rho^{\prime\prime}=\sum_{\alpha ,\beta}P_{\alpha ,\beta}\hat\rho_{\alpha ,\beta},\label{eq:simulrhoprime}
\end{equation}
where $P_{\alpha ,\beta}$ are suitable factors.  However, (\ref{eq:density}) does not take the form of (\ref{eq:simulrhoprime}).  Therefore, the states of $M_z$ and $M_x$ in (\ref{eq:phit0}) are entangled.

In weak measurement, the similar entanglement of the apparatuses is observed. 
If we weakly measure an observable $\hat A$ and then select the final state $|F\rangle$, the state of the unified system of the observed system and the two measuring devices, one of which weakly measures $\hat A$ and the other selects the final state, following the interaction between them is
\begin{equation}
|\Psi(t)\rangle=\exp(-i\hat H_F\Delta t )\exp(-i\hat H_A\Delta t )|\Psi(0)\rangle,\label{eq:enta}
\end{equation}
where $|\Psi(0)\rangle$ is the initial state of the combined system and $\hat H_F$ and $\hat H_A$ are defined as 
\[
\hat H_F=g_F|F\rangle\langle F|\hat\pi_F,
\]\[
\hat H_A=g_A\hat A\hat\pi_A,
\]
Hereafter, we put $g_F\Delta t=1$ for simplicity.  

We define the partial density matrix $\hat \rho_w^{(m)}(t)$ of the measuring devices as
\begin{equation}
\hat \rho_w^{(m)}(t)={\rm Tr}^{(s)}\big[|\Psi(t)\rangle\langle\Psi(t)|\big].\label{eq:densitypsi}
\end{equation}
By calculating the expectation value of either the position operator $\hat x_A$ of the probe of the measuring device of $\hat A$ or $\hat x_F$, we can obtain the expectation value of either $\hat A$ or $\hat F=|F\rangle\langle F|$ accurately as follows: 
\begin{equation}
\overline x_A= {\rm Tr}\big[\hat\rho_w^{(m)}(t)\hat x_A\big]
=g_A\Delta t\langle I|\hat A|I\rangle,
\end{equation}
\begin{equation}
\overline{x}_F= {\rm Tr}\big[\hat\rho_w^{(m)}(t)\hat x_F\big]=\langle I|\hat F|I\rangle.\label{eq:Fdake}
\end{equation}
Because of the reason which is almost the same as the previous discussion, we cannot know the expectation values of both $\hat A$ and $\hat F$ simultaneously, but we should regard the measurement of $\hat x_A$ and $\hat x_F$ as {\it one} manipulation.  The measured observable is $\hat x_F\hat x_A$, whose expectation value is
\begin{equation}
\begin{array}{rl}
\overline{x_Fx_A}&={\rm Tr}\big[ \hat x_F\hat x_A\hat\rho_w^{(m)}(t)\big]\\
&={\rm Tr}\big[ \hat x_A\hat x_F\hat\rho_w^{(m)}(t)\big]\\
&=\frac{1}{2}g_At_A\langle I|(\hat F\hat A+\hat A\hat F)|I\rangle.\label{eq:FAAF}
\end{array}
\end{equation}
Because we select cases of the measured value $X_F$ of $\hat x_F$, which is $1$ or $0$, is $1$ in the post-selection, the average of the measured value $X_A$ of $\hat x_A$ is equal to the average of $X_FX_A$ after the post-selection:
\begin{equation}
\langle X_A\rangle^{(p)}=\langle X_FX_A\rangle^{(p)},
\end{equation}
where $\langle\ \rangle^{(p)}$ stands for the average after post-selection. Without the post-selection, 
\begin{equation}
\langle X_FX_A\rangle=\overline{x_Fx_A},\label{eq:without}
\end{equation}
 Moreover, because $\langle X_FX_A\rangle^{(p)}$ is the quotient of the sum of post-selected $X_FX_A$'s, which is equal to the sum of all $X_FX_A$'s without any post-selection,  divided by the number of the post-selected data, it is boosted by $1/\langle X_F\rangle$:
\begin{equation}
\frac{\langle X_AX_F\rangle^{(p)}}{\langle X_AX_F\rangle}=\frac{1}{\langle X_F\rangle},
\end{equation}
where $\langle X_F\rangle$ is nearly equal to $\overline x_F$, because the first measurement is weak.   

Gathering these pieces, we obtain 
\begin{equation}
\langle X_A\rangle^{(p)}=\frac{\overline{x_Ax_F}}{\overline x_F}.\label{eq:average}
\end{equation}
By means of (\ref{eq:Fdake}) and (\ref{eq:FAAF}), (\ref{eq:average}) becomes
\begin{equation}
\frac{\langle X_A\rangle^{(p)}}{g_At_A}=\frac{\langle I|(\hat F\hat A+\hat A\hat F)|I\rangle}{2\langle I|\hat F|I\rangle}.\label{eq:real}
\end{equation}
The right-hand side of (\ref{eq:real}) is the real part of the weak value
\begin{equation}
\langle \hat A\rangle^w\equiv\frac{\langle F|\hat A|I\rangle}{\langle F|I\rangle}.\label{eq:wv}
\end{equation}
As confirmed by many experiments, the measured value of the weak measurement agrees with the corresponding weak value.  We theoretically showed it by taking account of the entanglement between the states of the two measuring devices.  If we ignored it and regarded (\ref{eq:wv}) as a some kind of expectation value of $\hat A$, we would suffer from the anomalous weak value problem. Therefore, we assert that weak value is evidence of the entanglement between the states of the two measuring devices in (\ref{eq:enta}), thus in J(\ref{eq:phit0}).

\end{document}